\documentclass[twoside,twocolumn,english]{revtex4-1}
\usepackage[T1]{fontenc}
\usepackage[latin9]{inputenc}
\usepackage{geometry}
\usepackage{color,soul}
\usepackage{amsmath}
\usepackage{amssymb}
\usepackage{enumerate}
\usepackage{graphicx}
\usepackage{mathbbol}
\usepackage{amsfonts}
\usepackage{ulem}



\usepackage{babel}

\setstcolor{red}

\begin{document}
\global\long\def\l{\lambda}%
\global\long\def\ints{\mathbb{Z}}%
\global\long\def\nat{\mathbb{N}}%
\global\long\def\re{\mathbb{R}}%
\global\long\def\com{\mathbb{C}}%
\global\long\def\dff{\triangleq}%
\global\long\def\df{\coloneqq}%
\global\long\def\del{\nabla}%
\global\long\def\cross{\times}%
\global\long\def\der#1#2{\frac{d#1}{d#2}}%
\global\long\def\bra#1{\left\langle #1\right|}%
\global\long\def\ket#1{\left|#1\right\rangle }%
\global\long\def\braket#1#2{\left\langle #1|#2\right\rangle }%
\global\long\def\ketbra#1#2{\left|#1\right\rangle \left\langle #2\right|}%
\global\long\def\paulix{\begin{pmatrix}0  &  1\\
 1  &  0 
\end{pmatrix}}%
\global\long\def\pauliy{\begin{pmatrix}0  &  -i\\
 i  &  0 
\end{pmatrix}}%
\global\long\def\pauliz{\begin{pmatrix}1  &  0\\
 0  &  -1 
\end{pmatrix}}%
\global\long\def\sinc{\mbox{sinc}}%
\global\long\def\ft{\mathcal{F}}%
\global\long\def\dg{\dagger}%
\global\long\def\bs#1{\boldsymbol{#1}}%
\global\long\def\norm#1{\left\Vert #1\right\Vert }%
\global\long\def\H{\mathcal{H}}%
\global\long\def\tens{\varotimes}%
\global\long\def\rationals{\mathbb{Q}}%
 
\global\long\def\tri{\triangle}%
\global\long\def\lap{\triangle}%
\global\long\def\e{\varepsilon}%
\global\long\def\broket#1#2#3{\bra{#1}#2\ket{#3}}%
\global\long\def\dv{\del\cdot}%
\global\long\def\eps{\epsilon}%
\global\long\def\rot{\vec{\del}\cross}%
\global\long\def\pd#1#2{\frac{\partial#1}{\partial#2}}%
\global\long\def\L{\mathcal{L}}%
\global\long\def\inf{\infty}%
\global\long\def\d{\delta}%
\global\long\def\I{\mathbb{I}}%
\global\long\def\D{\Delta}%
\global\long\def\r{\rho}%
\global\long\def\hb{\hbar}%
\global\long\def\s{\sigma}%
\global\long\def\t{\tau}%
\global\long\def\O{\Omega}%
\global\long\def\a{\alpha}%
\global\long\def\b{\beta}%
\global\long\def\th{\theta}%
\global\long\def\l{\lambda}%

\global\long\def\Z{\mathcal{Z}}%
\global\long\def\z{\zeta}%
\global\long\def\ord#1{\mathcal{O}\left(#1\right)}%
\global\long\def\ua{\uparrow}%
\global\long\def\da{\downarrow}%
 
\global\long\def\co#1{\left[#1\right)}%
\global\long\def\oc#1{\left(#1\right]}%
\global\long\def\tr{\mbox{tr}}%
\global\long\def\o{\omega}%
\global\long\def\nab{\del}%
\global\long\def\p{\psi}%
\global\long\def\pro{\propto}%
\global\long\def\vf{\varphi}%
\global\long\def\f{\phi}%
\global\long\def\mark#1#2{\underset{#2}{\underbrace{#1}}}%
\global\long\def\markup#1#2{\overset{#2}{\overbrace{#1}}}%
\global\long\def\ra{\rightarrow}%
\global\long\def\cd{\cdot}%
\global\long\def\v#1{\vec{#1}}%
\global\long\def\fd#1#2{\frac{\d#1}{\d#2}}%
\global\long\def\P{\Psi}%
\global\long\def\dem{\overset{\mbox{!}}{=}}%
\global\long\def\Lam{\Lambda}%
 
\global\long\def\m{\mu}%
\global\long\def\n{\nu}%

\global\long\def\ul#1{\underline{#1}}%
\global\long\def\at#1#2{\biggl|_{#1}^{#2}}%
\global\long\def\lra{\leftrightarrow}%
\global\long\def\var{\mbox{var}}%
\global\long\def\E{\mathcal{E}}%
\global\long\def\Op#1#2#3#4#5{#1_{#4#5}^{#2#3}}%
\global\long\def\up#1#2{\overset{#2}{#1}}%
\global\long\def\down#1#2{\underset{#2}{#1}}%
\global\long\def\lb{\biggl[}%
\global\long\def\rb{\biggl]}%
\global\long\def\RG{\mathfrak{R}_{b}}%
\global\long\def\g{\gamma}%
\global\long\def\Ra{\Rightarrow}%
\global\long\def\x{\xi}%
\global\long\def\c{\chi}%
\global\long\def\res{\mbox{Res}}%
\global\long\def\dif{\mathbf{d}}%
\global\long\def\dd{\mathbf{d}}%
\global\long\def\grad{\vec{\del}}%

\global\long\def\mat#1#2#3#4{\left(\begin{array}{cc}
#1 & #2\\
#3 & #4
\end{array}\right)}%
\global\long\def\col#1#2{\left(\begin{array}{c}
#1\\
#2
\end{array}\right)}%
\global\long\def\sl#1{\cancel{#1}}%
\global\long\def\row#1#2{\left(\begin{array}{cc}
#1 & ,#2\end{array}\right)}%
\global\long\def\roww#1#2#3{\left(\begin{array}{ccc}
#1 & ,#2 & ,#3\end{array}\right)}%
\global\long\def\rowww#1#2#3#4{\left(\begin{array}{cccc}
#1 & ,#2 & ,#3 & ,#4\end{array}\right)}%
\global\long\def\matt#1#2#3#4#5#6#7#8#9{\left(\begin{array}{ccc}
#1 & #2 & #3\\
#4 & #5 & #6\\
#7 & #8 & #9
\end{array}\right)}%
\global\long\def\su{\uparrow}%
\global\long\def\sd{\downarrow}%
\global\long\def\coll#1#2#3{\left(\begin{array}{c}
#1\\
#2\\
#3
\end{array}\right)}%
\global\long\def\h#1{\hat{#1}}%
\global\long\def\colll#1#2#3#4{\left(\begin{array}{c}
#1\\
#2\\
#3\\
#4
\end{array}\right)}%
\global\long\def\check{\checked}%
\global\long\def\v#1{\vec{#1}}%
\global\long\def\S{\Sigma}%
\global\long\def\F{\Phi}%
\global\long\def\M{\mathcal{M}}%
\global\long\def\G{\Gamma}%
\global\long\def\im{\mbox{Im}}%
\global\long\def\til#1{\tilde{#1}}%
\global\long\def\kb{k_{B}}%
\global\long\def\k{\kappa}%
\global\long\def\ph{\phi}%
\global\long\def\el{\ell}%
\global\long\def\en{\mathcal{N}}%
\global\long\def\asy{\cong}%
\global\long\def\sbl{\biggl[}%
\global\long\def\sbr{\biggl]}%
\global\long\def\cbl{\biggl\{}%
\global\long\def\cbr{\biggl\}}%
\global\long\def\hg#1#2{\mbox{ }_{#1}F_{#2}}%
\global\long\def\J{\mathcal{J}}%
\global\long\def\diag#1{\mbox{diag}\left[#1\right]}%
\global\long\def\sign#1{\mbox{sgn}\left[#1\right]}%
\global\long\def\T{\th}%
\global\long\def\rp{\reals^{+}}%

\title{Driven tracer dynamics in a one dimensional quiescent bath}
\author{Asaf Miron and David Mukamel}
\address{Department of Physics of Complex Systems, Weizmann
Institute of Science, Rehovot 7610001, Israel}

\begin{abstract}
The dynamics of a driven tracer in a quiescent bath subject to geometric confinement effectively models a broad range of phenomena. We explore this dynamics in a 1D lattice model where geometric confinement is tuned by varying particle overtaking rates.
Previous studies of the model's stationary properties on a ring of $L$ sites have revealed a phase in which the bath density profile extends over an $\sim \ord{L}$ distance from the tracer and the tracer's velocity vanishes as $\sim 1/L$. Here, we study the model's \textit{dynamics} in this phase as $L\ra \infty$ and for long times. We show that the bath density profile evolves on a $\sim \sqrt{t}$ time-scale and, correspondingly, that the tracer's velocity decays as $\sim 1/\sqrt{t}$. Unlike the well-studied non-driven tracer, whose dynamics becomes diffusive whenever overtaking is allowed, we here find that driving the tracer preserves its hallmark sub-diffusive single-file dynamics, even in the presence of overtaking.
\end{abstract}

\maketitle

\section{Introduction \label{sec:Introduction}}
The motion of a passive tracer, or tagged particle, in a bath of identical hard-core particles confined to a narrow environment models a broad range of phenomena. Such scenarios are abundant in biological systems, with examples including transport in
porins \cite{Nestorovich9789}, through the nuclear pores in eukaryotic cells \cite{rout2003virtual, kabachinski2015nuclear,wente2010nuclear} and along microtubules \cite{WELTE2004R525}.
In the case of a sufficiently narrow channel, where particles cannot overtake one another, the bath's correlated dynamics strongly restrict the tracer. Extensive studies of its motion in such settings, termed "single-file" (SF) dynamics, has led to the celebrated \textit{sub-diffusive} scaling $\sim\sqrt{t}$ of the tracer's mean-square displacement (MSD)
$\left\langle \D X\left(t\right)^{2}\right\rangle\equiv \left\langle X\left(t\right)^{2}\right\rangle - \left\langle  X\left(t\right)\right\rangle^{2}$ \cite{jepsen1965dynamics, percus1974anomalous, alexander1978diffusion}.
Evidently, this behavior becomes very fragile as the degree of geometric confinement is reduced to the point where particles can overtake one-another. In fact, it was shown that \textit{any} finite overtaking rate ultimately yields a diffusive scaling of the tracer's MSD, i.e. $\left\langle \D X\left(t\right)^{2}\right\rangle \sim t$ for large $t$  \cite{sane2010crossover,siems2012non,kumar2015crossover,ahmadi2017diffusion}.

Motivated by a variety of physical, biological and chemical setups \cite{grier2003revolution,wilson2011small,wittbracht2010flow}, as well as applications in microrheology \cite{squires2005simple}  and in nanotechnology and microfluidics \cite{kirby2010micro},
recent years have seen a growing interest in geometrically constrained systems in which the tracer is \textit{driven} or biased by an external force \cite{burlatsky1992directed,burlatsky1996motion,de1997dynamics,landim1998driven,benichou1999biased,illien2013active,cividini2016exact,cividini2016correlation,kundu2016exact,benichou2018tracer}.
Here too, when confinement is strong enough to prevent particle overtaking, the driven tracer's motion generates a blockade of bath particles that, in turn, restricts its propagation. This is manifested in the tracer's stationary velocity, which was shown to vanish as $v\sim L^{-1}$ on a 1D ring of length $L$, for large $L$ \cite{burlatsky1992directed,burlatsky1996motion,de1997dynamics,illien2013active,oshanin2004biased}.

The stationary behavior of a driven tracer model with \textit{finite} overtaking rates was recently studied on a ring of $L$ sites, occupied by $N$ symmetric hard-core bath particles of density $\overline{\r}=N/\left(L-1\right)$ \cite{miron2019single}.
Two distinct phases were identified and the model's phase diagram was determined in terms of the dynamical rates and $\bar{\r}$. Each of the two phases, called "localized" and "extended" respectively, was characterized by the corresponding stationary tracer velocity $v$ and bath particle density profile $\r_{\el}$, as seen in the tracer's frame of reference, where $\el$ denotes the site label. In the localized phase, the tracer attains a finite, $L$-independent velocity at large $L$ and the deviation of the bath density $\r_{\el}$ from its mean value $\overline{\r}$ remains localized around the tracer. On the other hand, in the extended phase the tracer's velocity vanishes as $\sim L^{-1}$ and the density profile $\r_{\el}$ continues to vary throughout the entire system. 

It is hard to avoid drawing a naive correspondence between the non-driven tracer's and the driven tracer's behaviors. The non-driven tracer's unrestricted diffusive behavior, which arises whenever overtaking is possible, is consistent with the finite velocity and localized bath density profile found in the driven tracer's localized phase. Similarly, the non-driven tracer's sub-diffusive dynamics, which appear in the absence of overtaking, is consistent with the vanishing velocity and extended bath density profile found in the driven tracer's extended phase. Yet, defying naive intuition, the extended phase was surprisingly shown in \cite{miron2019single} to persist for a range of finite overtaking rates.

In light of the intriguing stationary behavior found in the driven tracer's extended phase, one is left to wonder how are the model's \textit{dynamical} properties affected by overtaking. For example, how does the bath density's non-local profile evolve in time? How does the tracer's velocity vanish as $t\ra \infty$? Moreover, in the absence of overtaking, the sub-diffusive scaling which characterizes the non-driven tracer's MSD is well- known to extend to the driven case \cite{cividini2016correlation, driven_msd}. Can this behavior also persist in the presence of overtaking?

In this paper, we explore the \textit{dynamical} properties of a driven tracer, propagating in a crowded bath of symmetric bath particles subject to varying geometric confinement. This is carried out by studying the dynamics of the 1D lattice model introduced in \cite{miron2019single}, where geometric confinement is incorporated by allowing the driven tracer to overtake neighboring bath particles at fixed rates. Focusing on the model's "extended" phase, we use the mean-field approximation (MF) to compute the bath density profile's long-time asymptotic evolution, starting from the flat initial profile $\r_{\el}(0)=\overline{\r}$. This, in turn, is used to show that the tracer's velocity $v(t)$ decays as $\sim 1/\sqrt{t}$ for $t$ satisfying $1 \ll t \ll L^2$. We demonstrate, through extensive numerical simulations, that in this limit the classical result $\left\langle \D X\left(t\right)^{2}\right\rangle \pro\sqrt{t}$, which was obtained for the non-driven tracer in the absence of overtaking, remarkably persists in the driven tracer's extended phase, even in the presence of \textit{finite} overtaking rates. 
Finally, we observe that the tracer's dynamics can be reduced to that of a biased random walker with {\it{time-dependent hopping rates}}. Calculating these rates from the MF expression for the time-dependent bath density profile, allows us to recover the correct sub-diffusive scaling in a specific sub-region of this phase.

The paper is organized as follows: In Section II we introduce the
model. The main results are presented in Section III. In Section IV we carry out a MF analysis of the system's dynamics in the extended phase, which yields the temporal evolution of the bath density profile and the tracer's MD. These results are then used to construct the corresponding biased random walker dynamics and to calculate the tracer's MSD. In Section V we describe the numerical analysis. In Section VI concluding remarks are given.

\section{The Model\label{sec:The-Model}}

\begin{figure}
\begin{centering}
\includegraphics[scale=0.4]{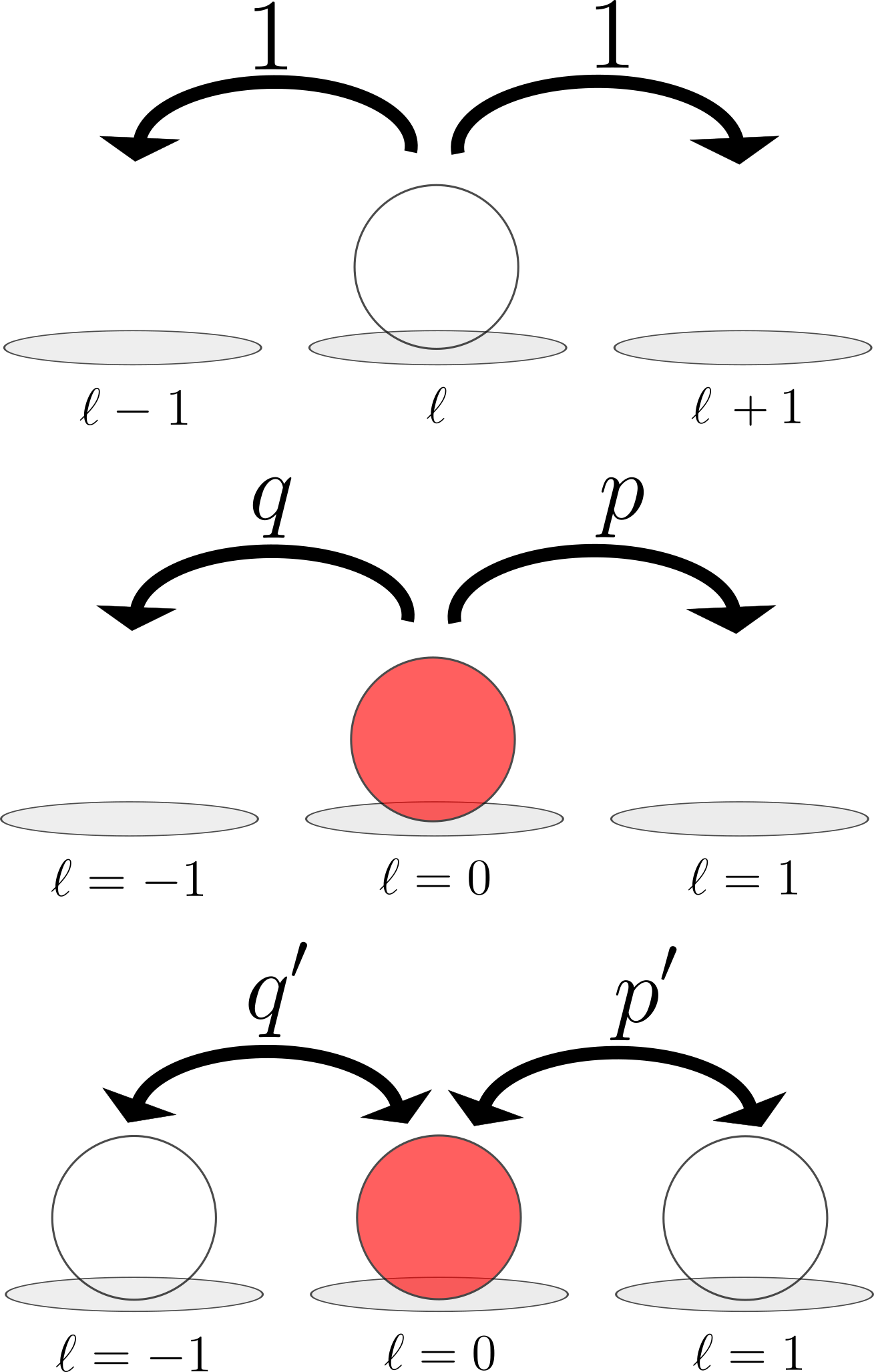} 
\par\end{centering}
\caption{Schematic illustration of the model dynamics. Bath particles are depicted by empty circles and the tracer is depicted by a red circle. Arrows represent the allowed moves, with their respective rates appearing above.}
\label{illustration} 
\end{figure}

Consider an infinite $1D$ ring with sites
$\el=-\infty,...,-2,-1,0,1,2,...,+\infty$. 
Working in the tracer's reference frame, site $\el=0$ is set to be the tracer's position at all times while, at $t=0$, the remaining sites are uniformly occupied by the bath particles whose average density is $\overline{\r}$. The particles interact via hard-core exclusion, whereby each site may be occupied by one particle, at most. The distinction between the bath particles and the tracer is manifested in their different dynamical rates: bath particles attempt to hop to a vacant neighboring site, on either side, with rate $1$ while the tracer attempts to hop to the right and left with respective rates $p$ and $q$. The varying degree of geometric confinement is incorporated by allowing the tracer to overtake, or exchange places with, a neighboring bath particle at rate $p'$ to the right and $q'$ to the left, if a bath particle is present there. This dynamics is schematically illustrated in Fig. \ref{illustration}.

The model's stationary behavior on a finite ring of $L$ sites was thoroughly explored in \cite{miron2019single}. Its phase diagram was obtained in the MF approximation and was shown to feature a non-equilibrium phase transition between a "localized" phase, where the tracer attains a finite velocity as $L\ra\infty$ and the bath density profile deviates from $\bar{\r}$ only in an $\sim\ord 1$ region around the tracer, and an "extended" phase, where the tracer's velocity vanishes as $\sim L^{-1}$ and the bath density profile extends over an $\sim\ord{L}$ region.

\begin{figure}
\begin{centering}
\includegraphics[scale=0.6]{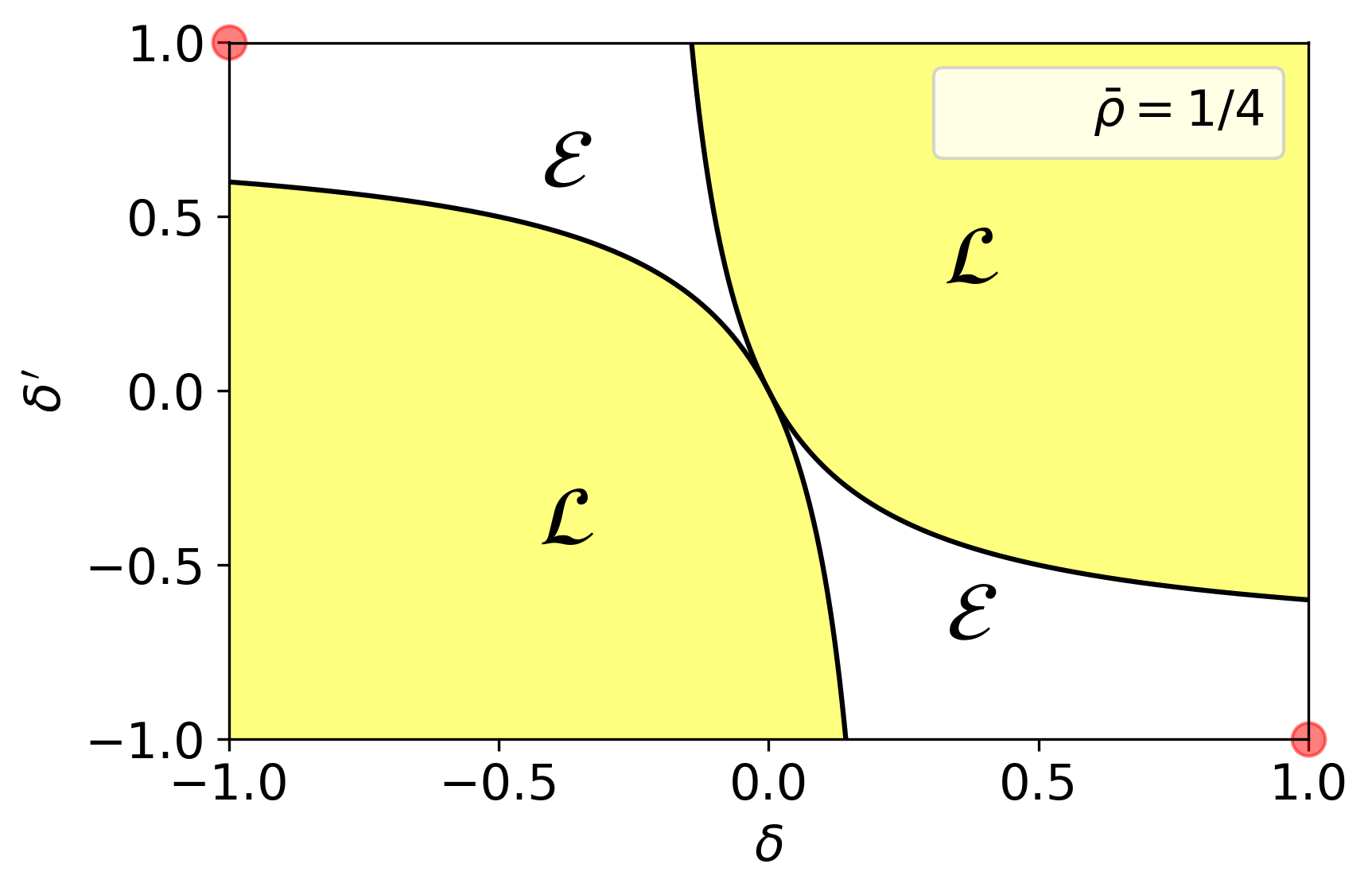} 
\par\end{centering}
\caption{The MF phase diagram for average bath density $\overline{\protect\r}=1/4$.
The localized and extended phases are respectively denoted by $\mathcal{L}$
and $\mathcal{E}$. Red circles mark the "extreme" points of the extended phase.}
\label{PD} 
\end{figure}

The model's phase diagram is most conveniently presented when the
hopping and exchange rates are rewritten as 
\begin{equation}
\begin{cases}
p=r\left(1+\delta\right)\text{ };\text{ }q=r\left(1-\delta\right)\\
p'=r'\left(1+\delta'\right)\text{ };\text{ }q'=r'\left(1-\delta'\right)
\end{cases},\label{eq: delta and delta prime}
\end{equation}
where $r\ge 0$ and $r'\ge 0$ are the respective average hopping and exchange rates, while $r\delta$ and $r'\delta'$ are the biases with $-1\le\delta,\delta'\le1$. Two critical manifolds, separating the extended and localized phases, were identified at the mean bath densities $\overline{\rho}_{c}^{I}$ and $\overline{\rho}_{c}^{II}$
\begin{equation}
\begin{cases}
\overline{\r}_{c}^{I}=\frac{q'\left(p-q\right)}{pq'-qp'}\equiv\frac{\delta\left(1-\delta'\right)}{\delta-\delta'}\\
\overline{\r}_{c}^{II}=\frac{p'\left(p-q\right)}{pq'-qp'}\equiv\frac{\delta\left(1+\delta'\right)}{\delta-\delta'}
\end{cases}.\label{eq:rho_c-1}
\end{equation}
Since these two manifolds are independent of the average rates $r$ and $r'$, the phase diagram may be represented in the 3D parameter space $\left\{\overline{\r},\delta,\delta'\right\}$. For convenience, and without loss of generality, we shall hereafter explicitly consider $\delta>0$ (i.e. $p>q$). The phase diagram in the $\left(\delta',\delta \right)$ plane is depicted in Fig. \ref{PD} for the mean density $\overline{\r}=1/4$.

The focus of this study lies on the model's dynamical properties in the extended phase. To this end, we distinguish between two sub-regions within the extended phase. The first region consists of the "extreme" points marked in red in Fig. \ref{PD}, i.e. $\d=-\d'=1$ and $\d=-\d'=-1$. For $\d=-\d'=1$ the dynamical rates are $q=p'=0$ and $p,q'>0$, implying that the hopping process is \textit{fully} biased to the right and the exchange process is \textit{fully} biased to the left. A symmetric and opposite picture arises for $\d=-\d'=-1$ where $p=q'=0$ and $q,p'>0$. The second region is the remaining "bulk" of the extended phase, where the hopping and exchange rates are only \textit{partially} counter-biased.

\section{Main Results\label{sec:Main-Results}}
The model's dynamical properties are studied in the extended phase using MF analysis and numerical simulations. 

Starting from a flat initial condition, the bath density profile at long times is shown in Eqs. \eqref{eq:ansatz} and \eqref{eq:density solution_1} to become a scaling function of the variables
\begin{equation}
z_{\pm}=\frac{c_{\pm}+\ell}{\sqrt{t}},
\label{eq:z scaling}
\end{equation}
where $z_+$ and $z_-$ correspond to $\ell>0$ and $\ell<0$, respectively.
The parameters $c_{\pm}$ are determined by the corresponding boundary conditions in each of the two regions. This result is then used to compute the tracer's velocity $v(t)$, which is shown in Eq. \eqref{eq:v(t)} to decay as $\sim 1/\sqrt{t}$ for large $t$.

The driven tracer's MSD in the extended phase is numerically shown to retain the classical sub-diffusive scaling $\left\langle \D X\left(t\right)^{2}\right\rangle \pro\sqrt{t}$, even for finite overtaking rates. This stands in contrast with the well-established results of numerous studies of non-driven tracer models, where diffusion arises for any finite overtaking rates. 

Remarkably, the sub-diffusive $\sim\sqrt{t}$ scaling of the MSD can be derived, at the extreme points $\d=-\d'=1$ and $-\d=\d'=1$,  \textit{within} the MF approximation. This is possible in-spite of the fact that different-time correlations are inherently absent from this description. To this end, we model the tracer's dynamics as a biased random walk with time-dependent rates. For "extreme" parameters, the bath density near the tracer satisfies $\rho_1 \approx 1-a/\sqrt{t}$ and $\rho_{-1} \approx b/\sqrt{t}$, where $a$ and $b$ may be read off Eq. \eqref{eq:density solution small l}. Thus, both the density of vacancies to the right of the tracer and the density of bath particles to its left approach $0$ as $\sim 1/\sqrt{t}$ in the long time limit. As such, due to the fully-biased nature of the dynamical rates, the tracer carries out right and left moves with rates which decrease in time as $\sim 1/\sqrt{t}$. The tracer's position is then modeled by a biased random walk with time-dependent move probabilities which decrease with time as $\sim 1/\sqrt{t}$. When correlations between consecutive moves are neglected, we obtain $\left\langle \D X\left(t\right)^{2}\right\rangle \pro\sqrt{t}$. Moreover, in a finite system of $L$ sites, we use this approach to establish the MSD's scaling behavior in two distinct dynamical regions: for $t\ll L^2$ the tracer's dynamics is sub-diffusive, as is the case for single file dynamics, while at $t\sim O(L^2)$ a crossover to ordinary diffusion takes place with a diffusion constant $D(L)$ which vanishes as $\sim 1/L$ at large $L$. 
While the description of the tracer's motion in terms of a random walk indeed captures the correct scaling behavior at the extreme points, we note that this approach fails in the bulk of the extended phase, where the bath density near the tracer reaches neither $0$ nor $1$ and correlations cannot be safely discarded.

\begin{figure*}
\centering{}\includegraphics[scale=0.575]{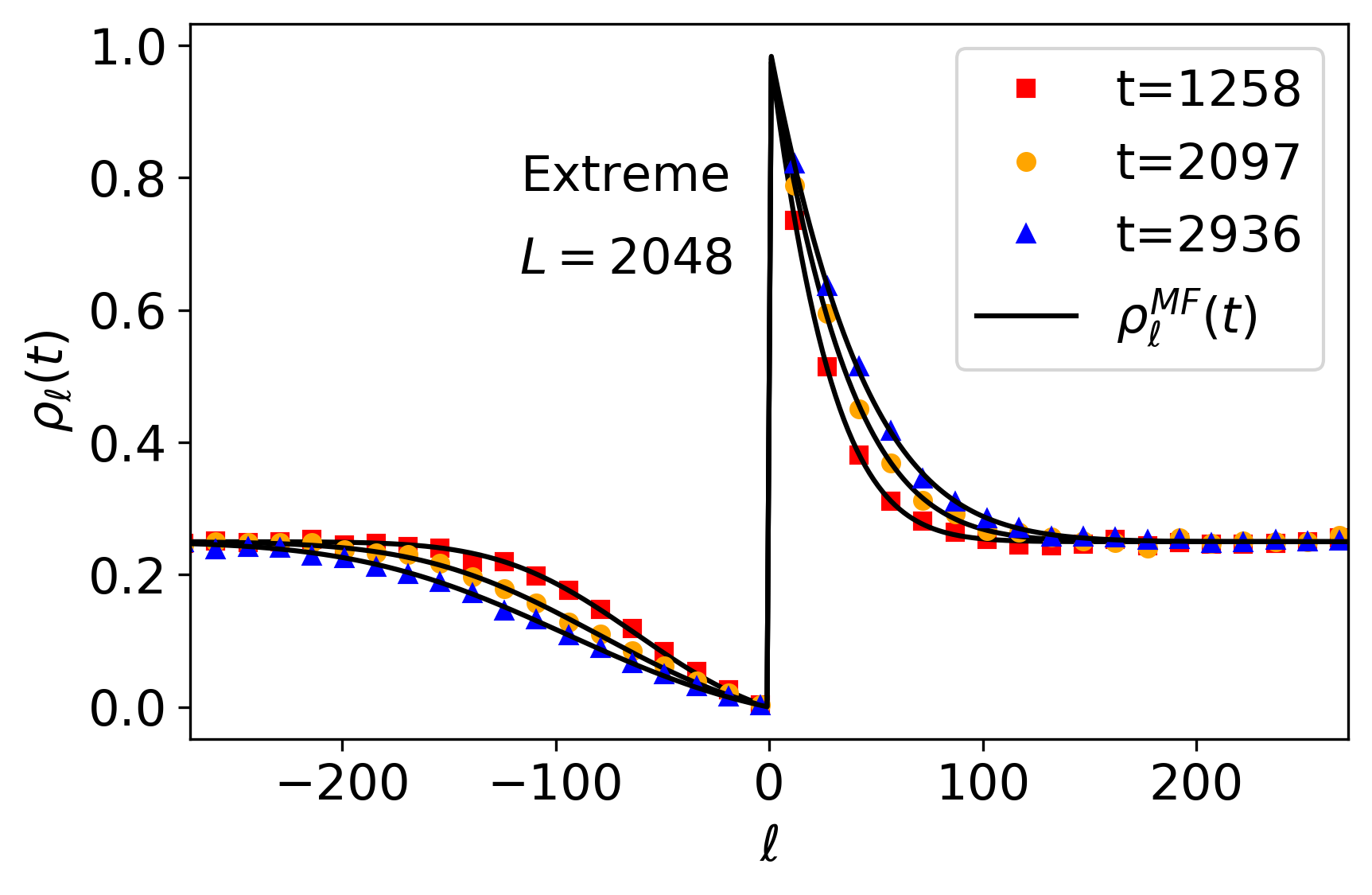}\hfill{}
\includegraphics[scale=0.6]{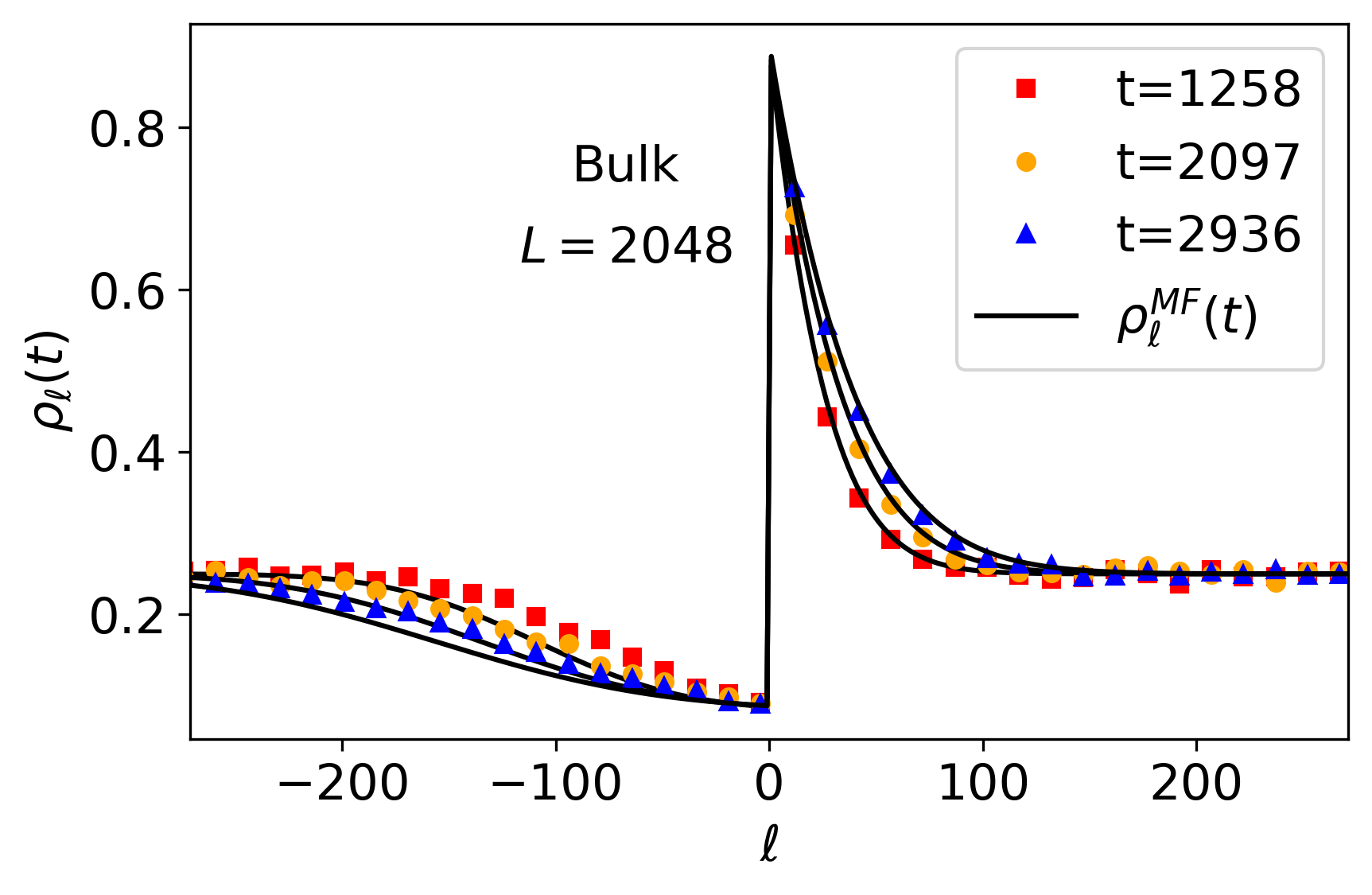}\caption{The density profile $\rho_{\ell}(t)$ versus $\ell$ in a region of $540$ sites around the tracer. Different times are denoted by different markers and MF predictions are depicted by black lines.}
\label{densities_ell} 
\end{figure*}

\begin{figure*}
\centering{}\includegraphics[scale=0.575]{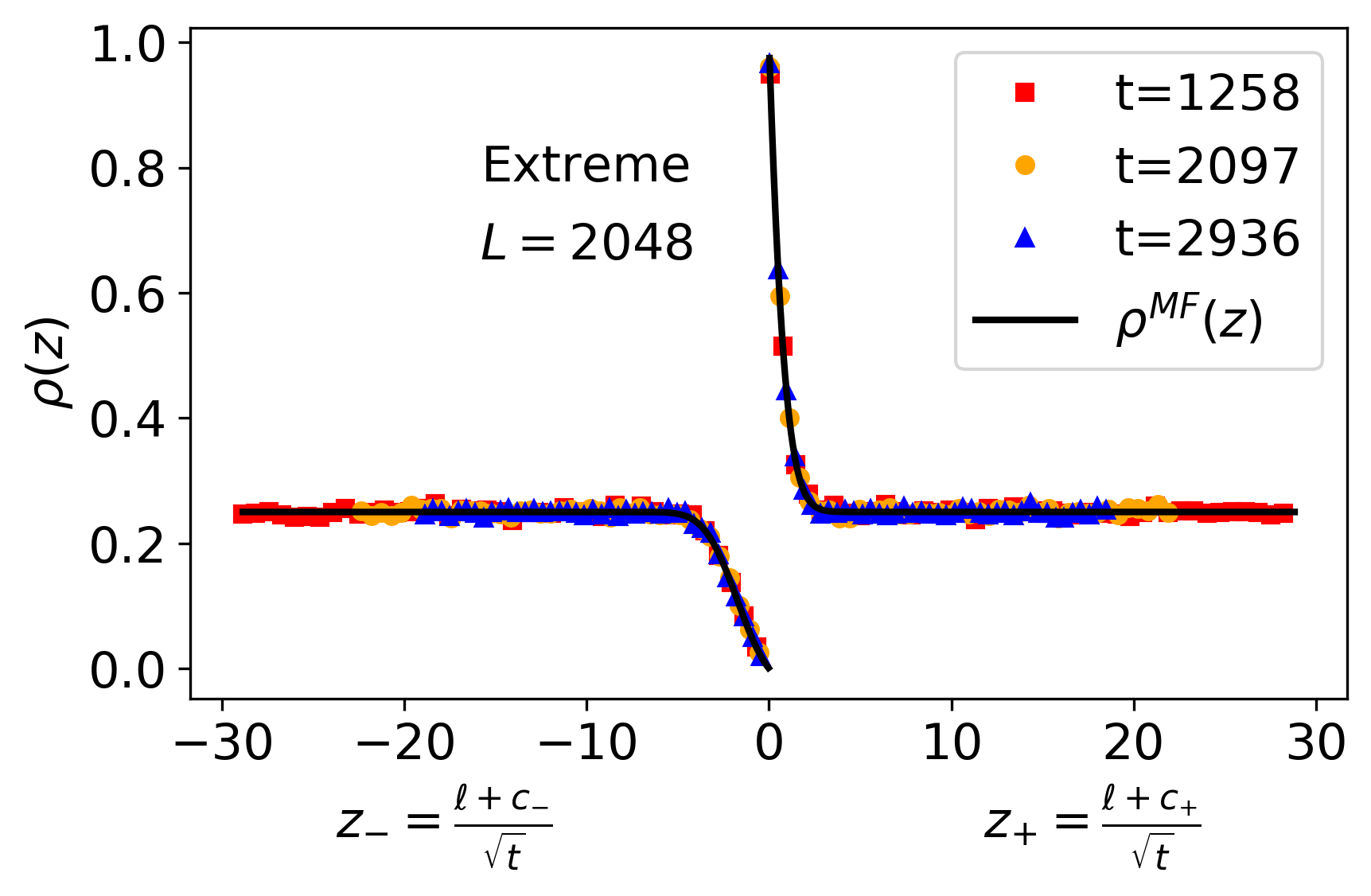}\hfill{}
\includegraphics[scale=0.6]{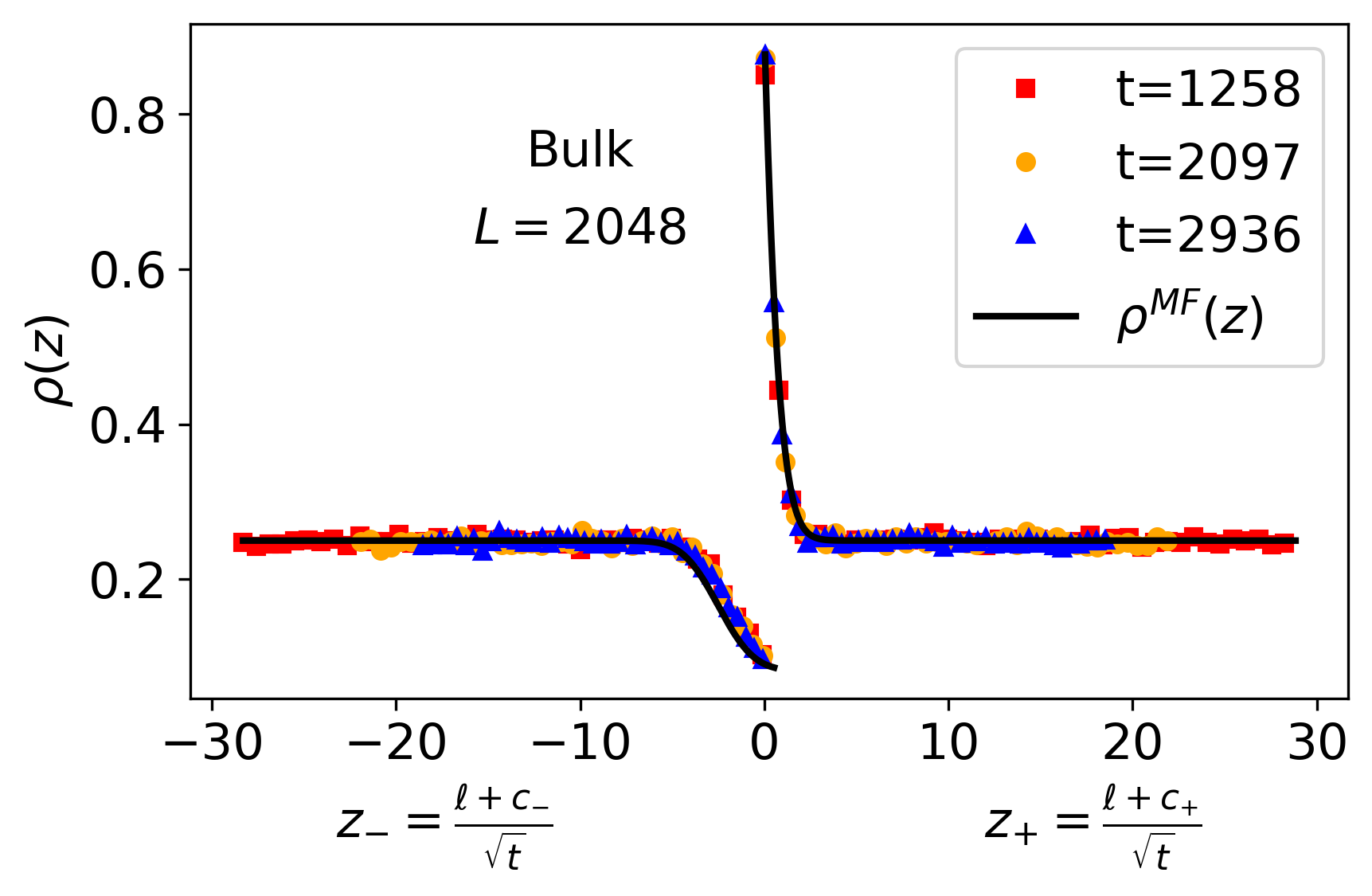}\caption{The density profile $\rho(z)$ versus the scaling variable $z$ for different times. The different times are denoted by different markers and MF predictions are depicted by black lines.
}
\label{densities_z} 
\end{figure*}

\begin{figure}
\begin{centering}
\includegraphics[scale=0.58]{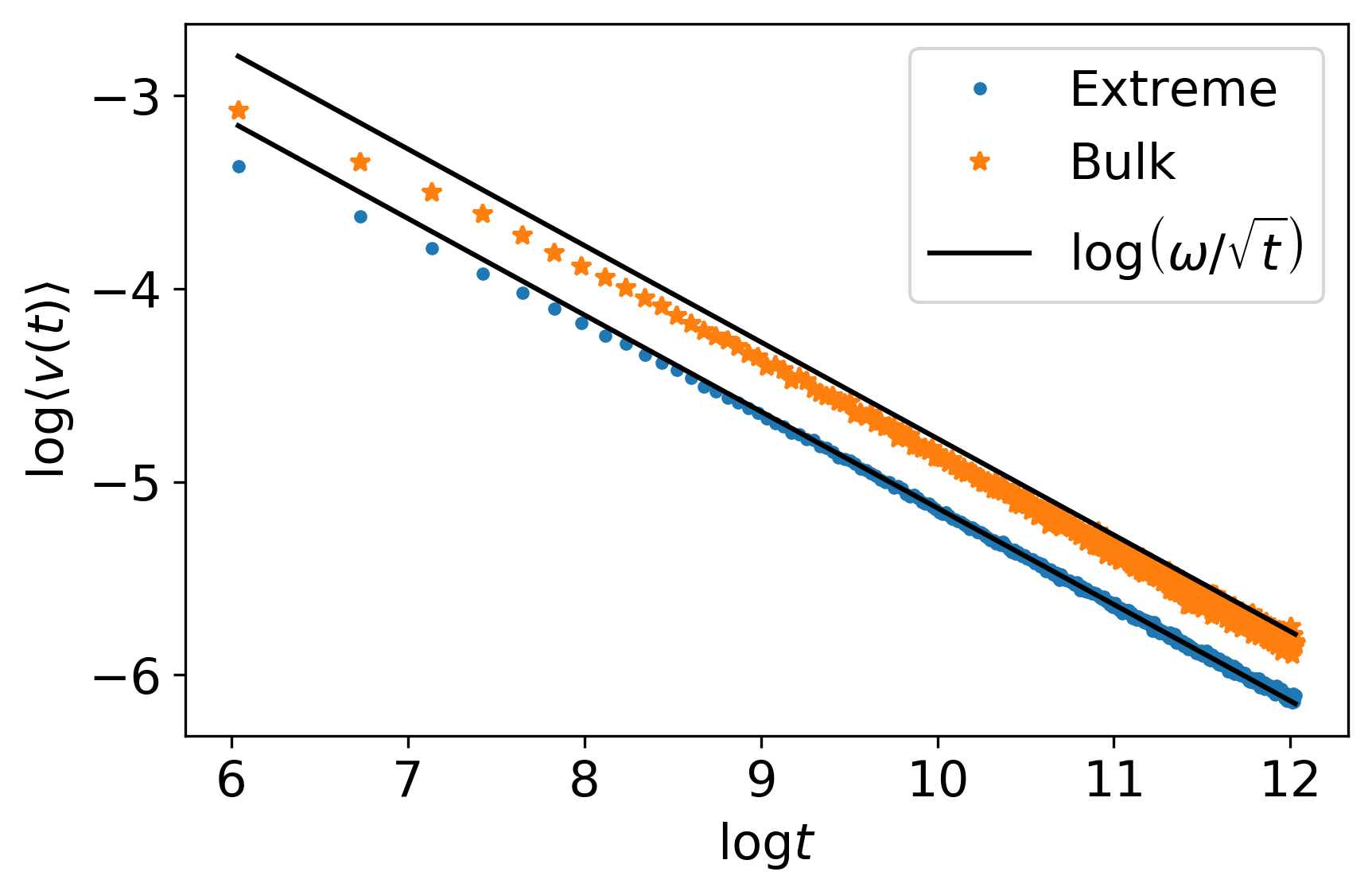} 
\par\end{centering}
\caption{Natural log-scale plot of the tracer's velocity versus time. Simulation results for the extreme parameters are denoted by blue dots while the bulk parameter results are denoted by orange stars. MF predictions are depicted by solid lines.}
\label{v} 
\end{figure}

\begin{figure*}
\centering{}\includegraphics[scale=0.575]{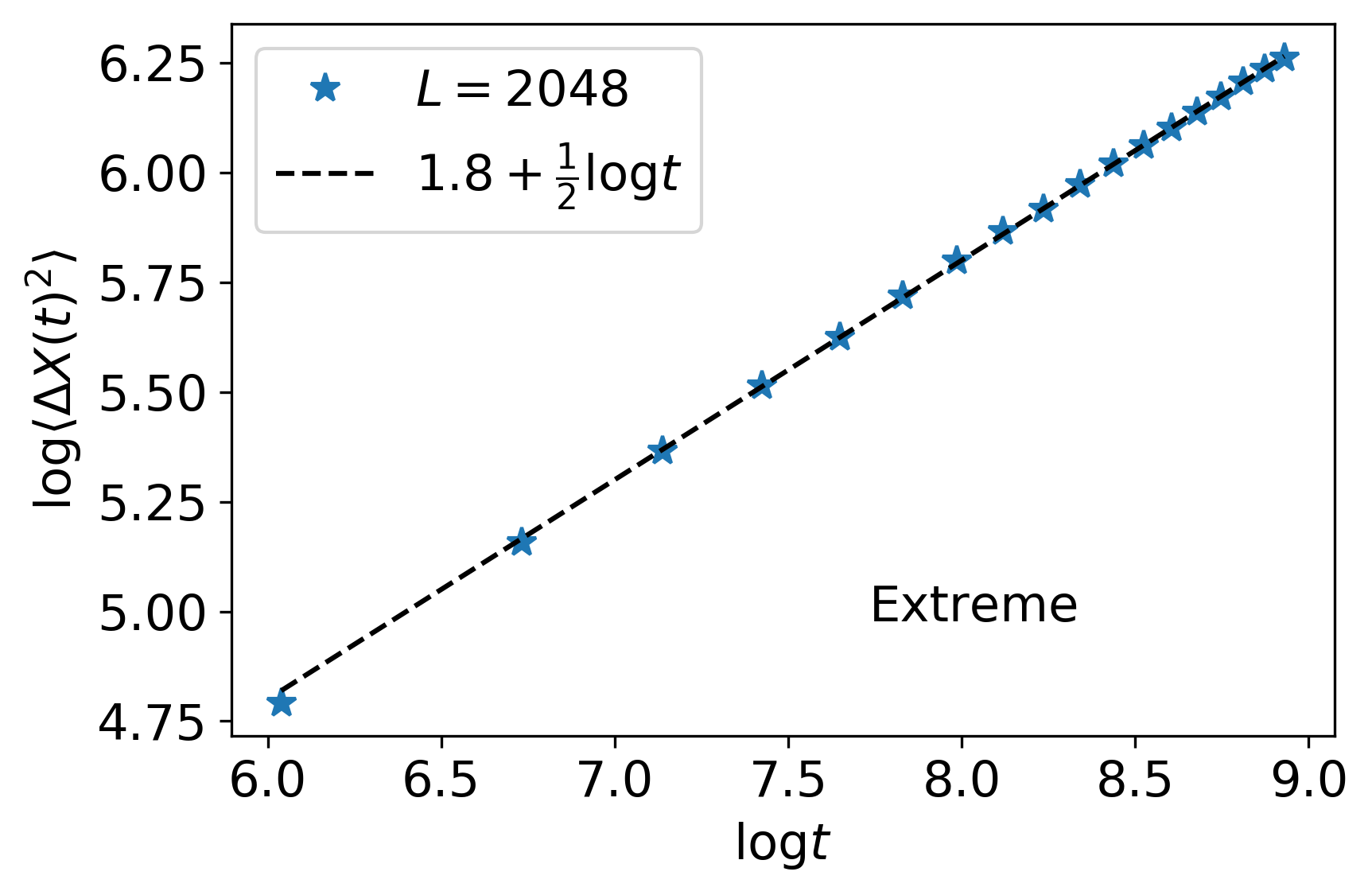}\hfill{}
\includegraphics[scale=0.6]{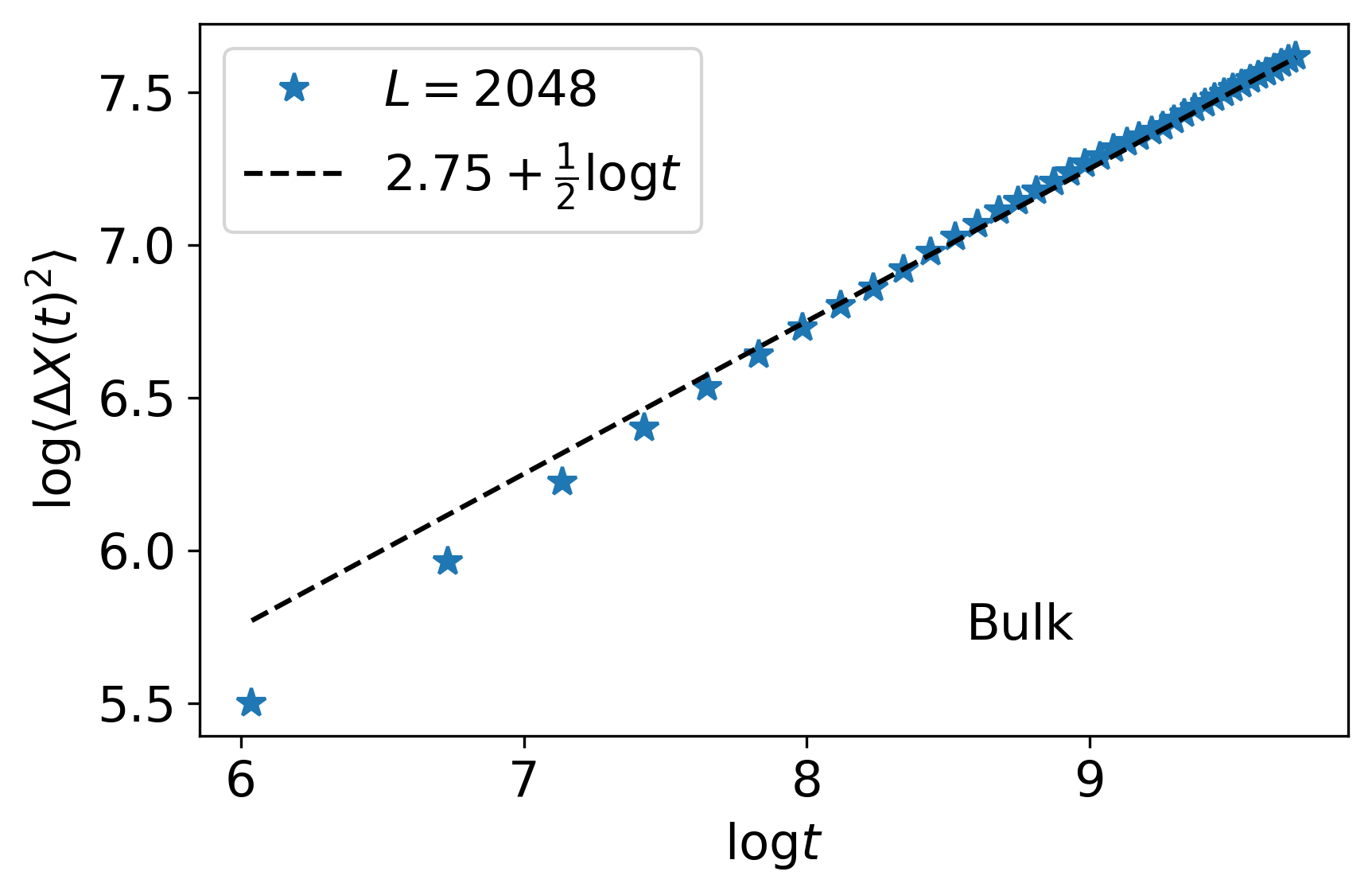}\caption{Log-log plot of the tracer's MSD versus time for short times. Blue stars depict simulation results for $L=2048$ and the dashed black curves depict a linear fit to $const+\frac{1}{2}\log{t}$. The left panel shows simulation results for the extreme parameters while the right panel shows results for the bulk parameters.}
\label{MSD_zoom} 
\end{figure*}

The following figures provide firm support of these results, presenting direct simulation results alongside
MF predictions. The simulation data is obtained from $\sim 10,000$ realizations of the model's dynamics for each set of parameters (i.e. rates and system sizes) with a fixed mean bath density of $\bar{\rho}=1/4$. The results depicted in Figs. \ref{densities_ell}, \ref{densities_z}, \ref{v} and \ref{MSD_zoom} are obtained in the long-time limit, but for times that are much shorter than the diffusive time scale with respect to the system size $L$, i.e. for $0 \ll t\ll L^{2}$. The figures that present results for the extreme region are obtained for the extreme rates $q=p'=0$ and $p=q'=1$ while results for the bulk region are obtained for the bulk rates $p=q'=1$ and $q=p'=0.1$. Both choices are appropriate representatives of the behavior found in the respective sub-regions of the extended phase.

In Fig. \ref{densities_ell} we plot the bath density profile $\r_{\ell}(t)$ near the tracer at different times. The left panel shows the extreme parameter results while the right panel shows the bulk parameter results. In both cases the profile appears to approach the mean bath density $\bar{\r}=1/4$ far from the tracer. An excellent fit to the MF prediction in Eq. \eqref{eq:density solution_1} is noted in the extreme region while a reasonable fit, slightly affected by correlations, is found for the bulk region. Figure \ref{densities_z} shows a data collapse of the density as a function of the scaling variables $z_{\pm}$ of Eq. \eqref{eq:ansatz}. Note that the collapse is equally convincing for both the bulk and extreme parameters, suggesting that the correlations observed in the bulk do not qualitatively change the density profile's scaling form.

We next show results for the tracer's dynamical properties, namely its velocity and MSD. 
The tracer's velocity $v(t)$ is presented in Fig. \ref{v} in natural log-log scale and simulation results for both the extreme and bulk regions are provided alongside their respective MF predictions, showing an excellent fit. 

\begin{figure*}
\centering{}\includegraphics[scale=0.575]{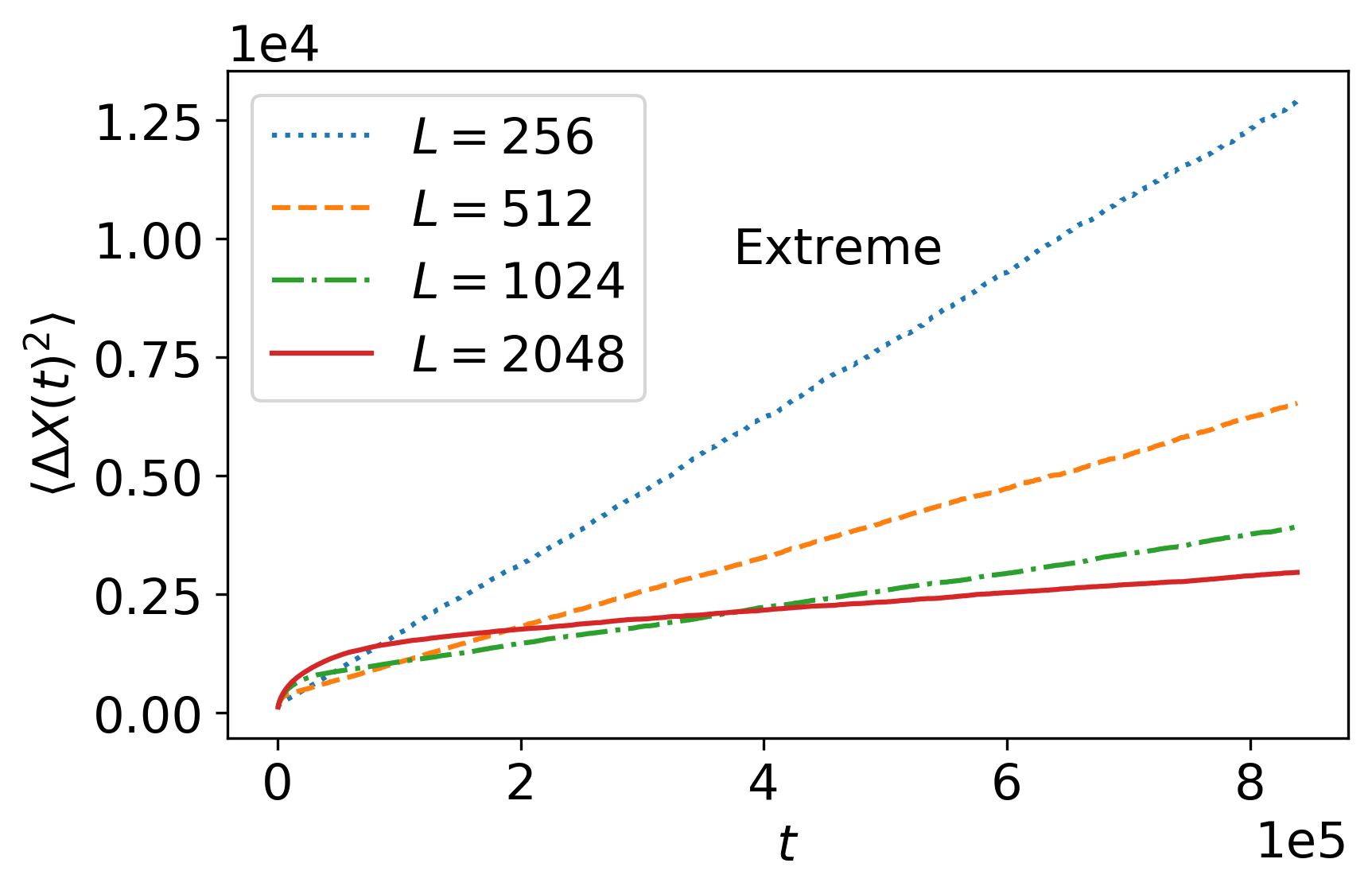}\hfill{}
\includegraphics[scale=0.6]{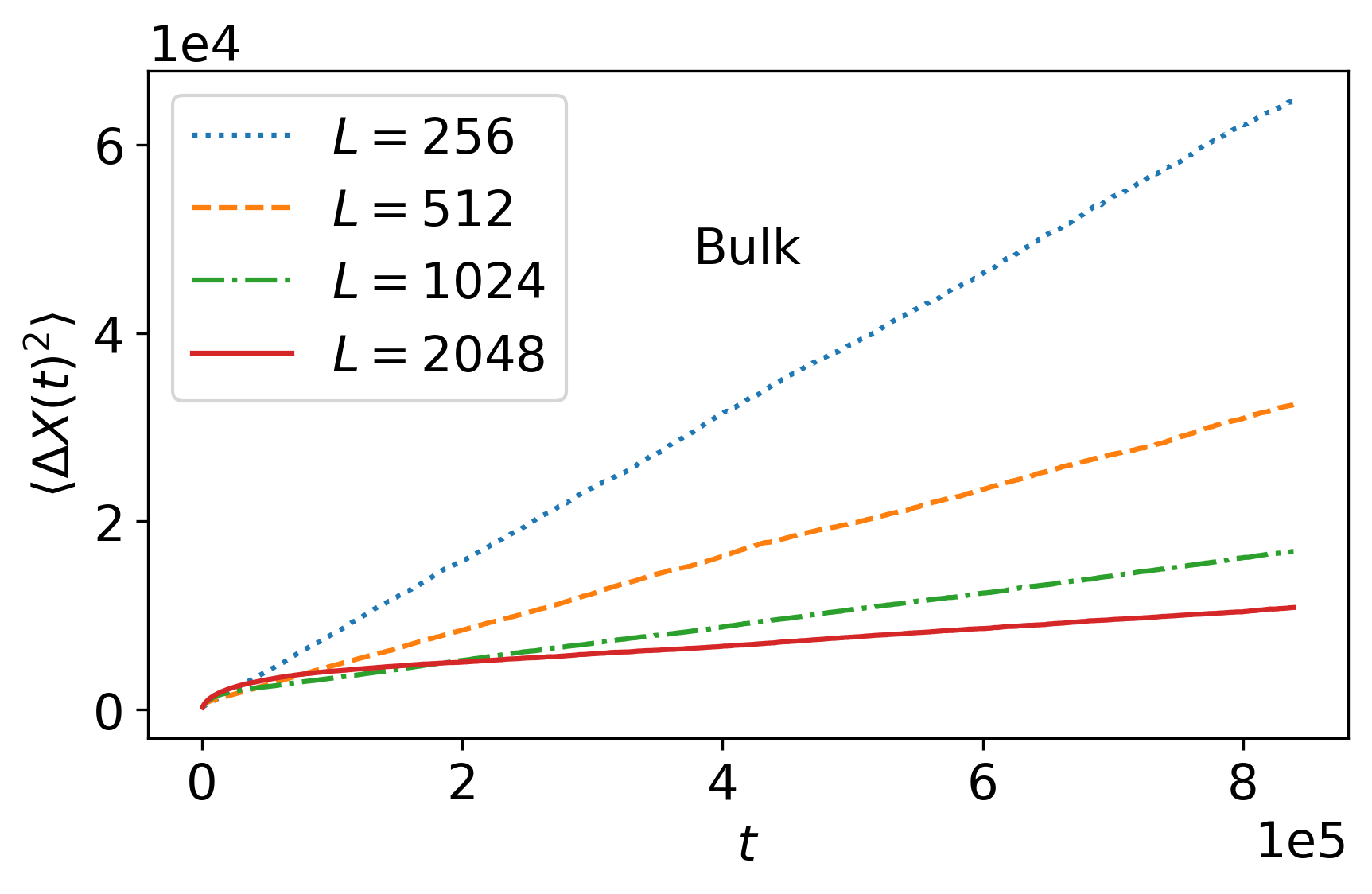}\caption{Plot of the tracer's MSD versus time for different values of $L$. The left panel shows simulation results for the extreme parameters while the right panel shows results for the bulk parameters.}
\label{MSD} 
\end{figure*}

\begin{figure}
\begin{centering}
\includegraphics[scale=0.58]{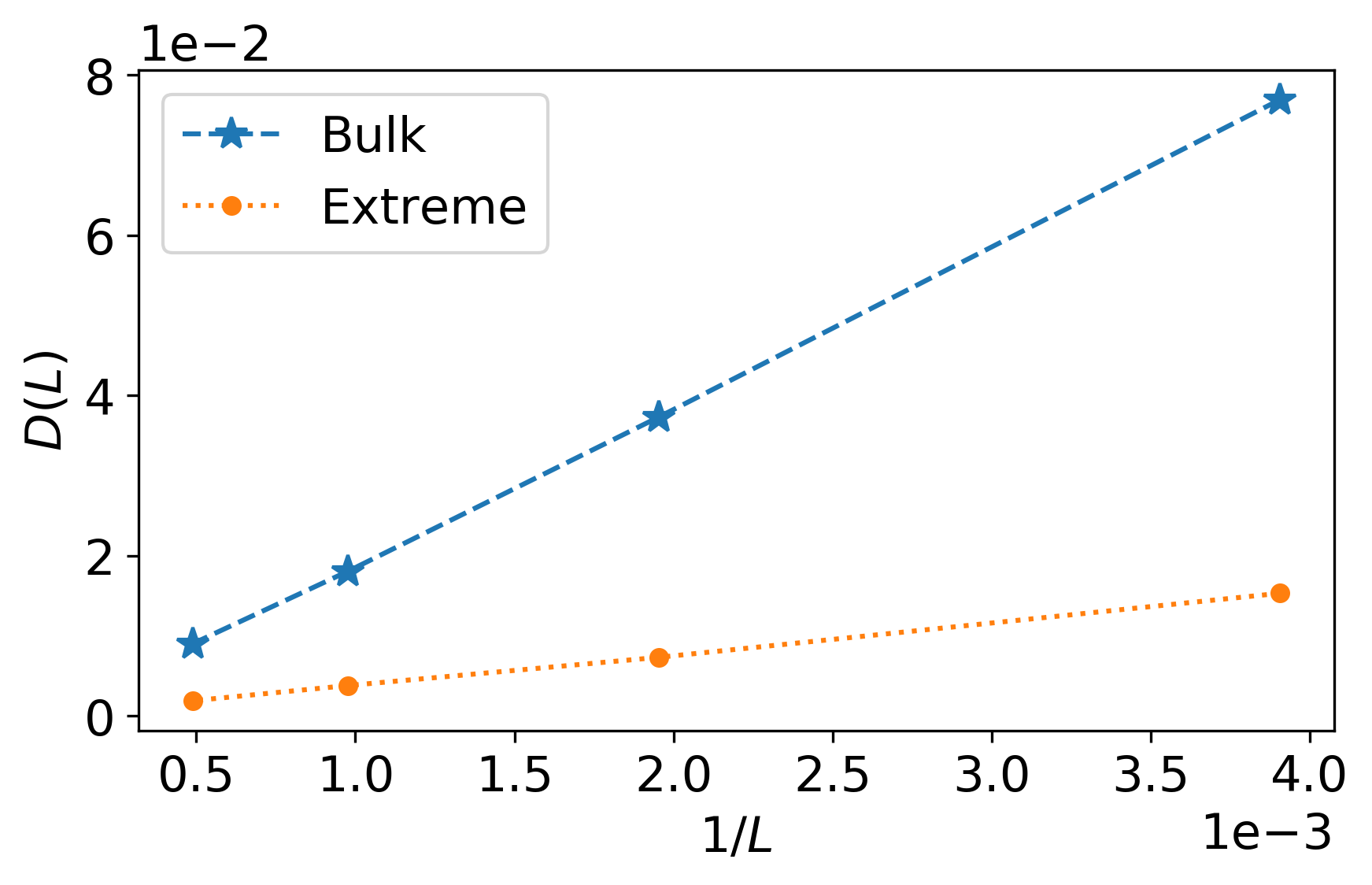} 
\par\end{centering}
\caption{The diffusion constant versus $1/L$ at long times satisfying $L^2 \ll t \ll L^3$. Blue stars show simulation data in the bulk while the orange dots depict the extreme points. The dashed and dotted curves simply serve as a guide for the eye.}
\label{MSD_slope} 
\end{figure}

\begin{figure*}
\centering{}\includegraphics[scale=0.575]{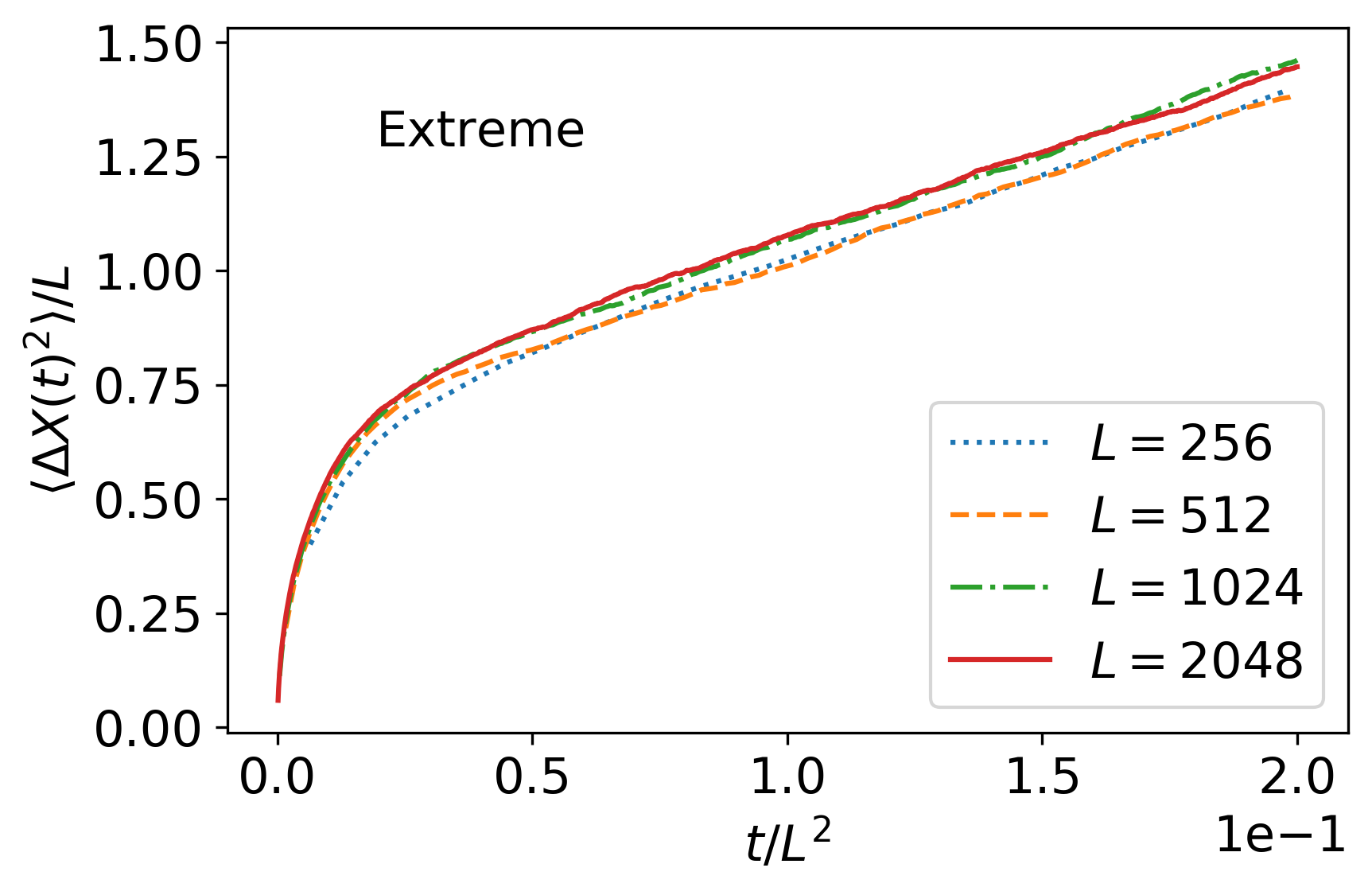}\hfill{}
\includegraphics[scale=0.6]{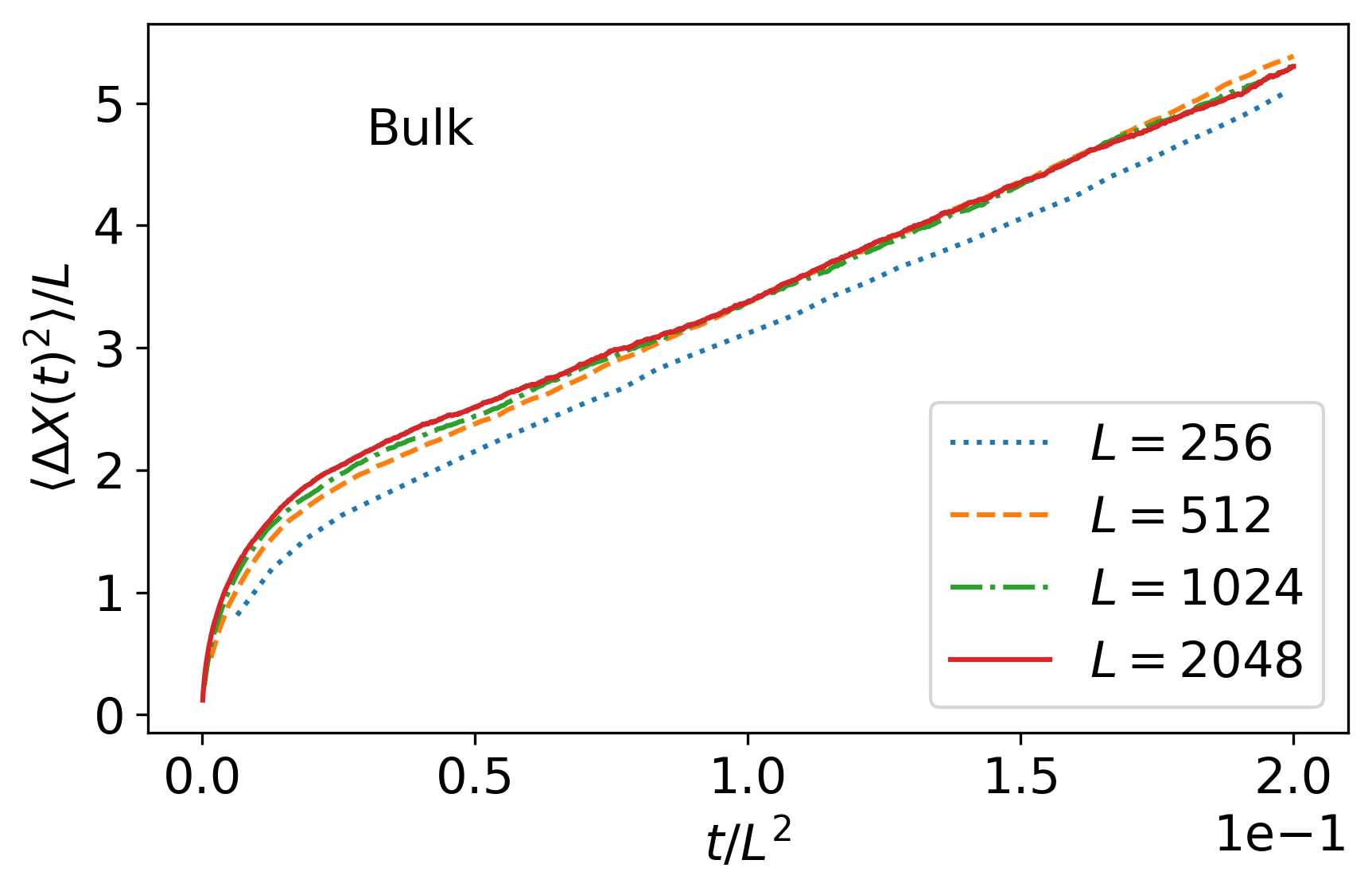}\caption{Collapse of the MSD scaled by the system size $L$ versus the diffusive timescale $t/L^2$ at long times and for different values of $L$.}
\label{MSD_scaled} 
\end{figure*}

We next consider the tracer's MSD, as obtained from numerical simulations.
In Fig. \ref{MSD_zoom} we zoom-in on the short-time behavior of the tracer's MSD in a large system of size $L=2048$. This figure shows that, for large times satisfying $0 \ll t\ll L^{2}$, the MSD scales as $\sim\sqrt{t}$ in both the extreme and bulk regions. 
In the long time limit, $t \gg L^2$ (but with $t\ll L^3$) the system reaches a steady state.
It is shown in Figs. \ref{MSD} and \ref{MSD_slope} that in this limit, the dynamics is diffusive with $\left\langle \D X\left(t\right)^{2}\right\rangle \simeq D(L)t$, where the diffusion coefficient $D(L)$ is found to decay as $\sim 1/L$ in the large $L$ limit at both the extreme and bulk regions of the phase diagram. The vanishing of $D(L)$ in the limit $L\ra\infty$ is consistent with the observed sub-diffusive $\sim\sqrt{t}$ scaling of the MSD for $t \ll L^2$. Figure \ref{MSD_scaled} further supports this picture, providing a data collapse of $\left\langle\Delta x(t)^{2}\right\rangle/L$ versus the scaling variable $t/L^{2}$ for different values of $L$.
Note that at even larger $t\sim O(L^3)$, another crossover takes place whereby the MSD stops growing with time, reaching its maximal value of $\sim O(L^2)$, as imposed by the system's finite size. 

\section{Dynamics\label{sec:Dynamics}}

Our analysis begins with formulating the equations which describe the
evolution of the bath occupation variable $\t_{\el}\left(t\right)$, which takes the value $1$ if site $\el$ is occupied at time $t$ and $0$ otherwise. Analyzing the contribution of each possible
process to $\t_{\el}\left(t\right)$ yields the following equations. Let 
\begin{equation}
\t_{\el}\left(t+dt\right)-\t_{\el}\left(t\right)=\G_{\el}\left(t\right). \label{eq:general tau}
\end{equation}
At the boundary sites $\ell=\pm 1$, $\G(t)$ is given by 
\begin{equation}
\G_{1}\left(t\right)=\begin{cases}
\t_{2} & \text{w.p. }\left(1+p\right)\left(1-\t_{1}\right)dt\\
-\t_{1} & \text{w.p. }\left[\left(1+p'\right)\left(1-\t_{2}\right)+q\left(1-\t_{-1}\right)\right]dt\\
\t_{-1} & \text{w.p. }q'\left(1-\t_{1}\right)dt
\end{cases}
,\label{eq:1 Gamma}
\end{equation}
and 
\begin{equation}
\G_{-1}\left(t\right)=\begin{cases}
\t_{-2} & \text{w.p. }\left(1+q\right)\left(1-\t_{-1}\right)dt\\
-\t_{-1} & \text{w.p. }\left[\left(1+q'\right)\left(1-\t_{-2}\right)+p\left(1-\t_{1}\right)\right]dt\\
\t_{1} & \text{w.p. }p'\left(1-\t_{-1}\right)dt
\end{cases}
,\label{eq:-1 Gamma}
\end{equation}
where "w.p." abbreviates with probability.
For the remaining sites $\el=-\infty,...,-2,2,...,\infty$, $\G(t)$ is given by
\begin{equation}
\G_{\el}\left(t\right)=\begin{cases}
\t_{\el+1} & \text{w.p. }\left(1-\t_{\el}\right)dt\\
\t_{\el-1} & \text{w.p. }\left(1-\t_{\el}\right)dt\\
-\t_{\el} & \text{w.p. }\left(1-\t_{\el\pm 1}\right)dt\\
\t_{\el+1}-\t_{\el} & \text{w.p. }\left[p\left(1-\t_{1}\right)+p'\t_{1}\right]dt\\
\t_{\el-1}-\t_{\el} & \text{w.p. }\left[q\left(1-\t_{-1}\right)+q'\t_{-1}\right]dt
\end{cases}.\label{eq:bulk Gamma}
\end{equation}

\subsection{MF approximation\label{subsec:MF-Approximation}}

The mean-field equations for the bath density profile $\r_{\el}\left(t\right)$ at
site $\el$ and time $t$ are obtained by averaging Eqs. \eqref{eq:general tau}, \eqref{eq:1 Gamma}, \eqref{eq:-1 Gamma} and \eqref{eq:bulk Gamma} over different realizations of the dynamics, such that $\r_{\el}\left(t\right)=\left\langle \t_{\el}\left(t\right)\right\rangle$. In the MF approximation, correlations between the occupation of two different sites are neglected, thus replacing terms of the form $\left\langle \t_{m}\t_{n}\right\rangle $ by the product $\left\langle \t_{m}\right\rangle \left\langle \t_{n}\right\rangle \ra\r_{m}\r_{n}$.
The boundary equations at sites $\el=\pm 1$ become 
\begin{equation}
\begin{cases}
\partial_{t}\r_{1}=\left(1-\r_{1}\right)\left[q'\r_{-1}+\left(1+p\right)\r_{2}\right]\\
-\left(1+p'\right)\r_{1}\left(1-\r_{2}\right)-q\r_{1}\left(1-\r_{-1}\right)\\
\partial_{t}\r_{-1}=\left(1-\r_{-1}\right)\left[p'\r_{1}+\left(1+q\right)\r_{-2}\right]\\
-\left(1+q'\right)\r_{-1}\left(1-\r_{-2}\right)-p\r_{-1}\left(1-\r_{1}\right)
\end{cases},\label{eq:boundary equations}
\end{equation}
and the general equation for $\el\ne\pm 1$ becomes
\begin{equation}
\partial_{t}\r_{\el}=\r_{\el+1}-2\r_{\el}+\r_{\el-1}+v_{+}\left(\r_{\el+1}-\r_{\el}\right)-v_{-}\left(\r_{\el}-\r_{\el-1}\right),\label{eq:rho_bulk_eq}
\end{equation}
where 
\begin{equation}
v_{+}=p\left(1-\r_{1}\right)+p'\r_{1}\label{eq:v_p}
\end{equation}
is the rate at which the tracer moves (either by hopping or exchange) to the right, while 
\begin{equation}
v_{-}=q\left(1-\r_{-1}\right)+q'\r_{-1}\label{eq:v_m}
\end{equation}
is the rate at which it similarly moves to the left. Note that, unlike the occupation variable $\t_{\el}$, the density profile $\r_{\el}$ takes continuous values in $\left[0,1\right]$. 

\subsubsection{Asymptotic scaling form}

Since the tracer permanently occupies site $\el=0$, the bath density profile
$\r_{\el}\left(t\right)$ is expected to be discontinuous at $\el=0$.
As such, we must separately treat Eqs. \eqref{eq:boundary equations} and \eqref{eq:rho_bulk_eq} for $\r_{\el}\left(t\right)$ in each of the two domains $\el=-\infty,...,-2,-1$ and $\el=1,2,...,\infty$. We denote the density profile in each of these respective regimes by $\r_{\el}^{-}\left(t\right)$ and $\r_{\el}^{+}\left(t\right)$.
We are interested in studying the long-time behavior of $\r_{\el}^{\pm}\left(t\right)$, starting from an initially flat density profile $\r_{\el}\left(0\right)=\overline{\r}$. Given that Eq. \eqref{eq:rho_bulk_eq} for the general dynamics at $\el\ne\pm1$ is simply a diffusion equation with drive, it is natural to seek a solution where $\el$ scales as $\sim\sqrt{t}$. In particular, we consider a large-time ansatz of the form 
\begin{equation}
\r_{\el}^{\pm}\left(t\right)=\overline{\r}+\d\r_{\pm}\left(z_{\pm}\right)\text{ };\text{ }z_{\pm}=\frac{\el+c_{\pm}}{\sqrt{t}},\label{eq:ansatz}
\end{equation}
where $\d\r_{\pm}\left(z_{\pm}\right)$ is the deviation from the
initially uniform density profile $\overline{\r}$ and $c_{\pm}$ are
yet-unknown parameters. These parameters may be neglected in the scaling regime where $|\el|$ is large and scales as $\sim\sqrt{t}$, but they \textit{must} be considered when solving the boundary equations for $\el=\pm 1$. At long times, $z_{\pm}$ approaches a continuous variable, such that $z_{-}\in\left(-\infty,0\right)$ and $z_{+}\in\left(0,+\infty\right)$.
In this limit, Eq. \eqref{eq:rho_bulk_eq} takes the continuous
form
\begin{equation}
0=\left(1+\frac{u}{2}\right)\ddot{\d\r_{\pm}}\left(z_{\pm}\right)+\left(v\sqrt{t}+\frac{z_{\pm}}{2}\right)\dot{\d\r_{\pm}}\left(z_{\pm}\right),\label{eq:rho_bulk_eq_scaled}
\end{equation}
where we define the respective tracer velocity $v\left(t\right)$ and moving rate $u\left(t\right)$ as 
\begin{equation}
v(t)=v_{+}(t)-v_{-}(t)\text{ },\text{ }u(t)=v_{+}(t)+v_{-}(t).\label{eq:v and u}
\end{equation}

\subsubsection{Velocity and moving rate}

In \cite{miron2019single}, the model's stationary behavior was studied on a finite chain of $L$ sites and the stationary density profile $\r_{\el}^{st}$ was recovered in the limit of large $L$. In the extended phase, where SF-like dynamical behavior might be expected, it was shown that
\begin{equation}
\r_{1}^{st}\approx \frac{q'\left(p-q\right)}{pq'-qp'}\text{ }\text{and}\text{ }\r_{L-1}^{st}=\frac{p'\left(p-q\right)}{pq'-qp'},\label{eq:extended phase boundary densities}
\end{equation} 
where corrections of $\sim\ord{L^{-1}}$ are neglected.
In our current scenario of an infinite chain where $\r_{L-1}^{st}\ra\r_{-1}^{st}$, one can show that substituting $\rho_{\pm1}^{st}$ into Eqs. \eqref{eq:v_p}, \eqref{eq:v_m} and \eqref{eq:v and u} yields a stationary tracer velocity and moving rate that behave as 
\begin{equation}
\lim_{t\ra\infty}v\left(t\right)\ra0\text{ }\text{and}\text{ }\lim_{t\ra\infty}u\left(t\right)\ra u_{0},\label{eq:v and u asymptotic}
\end{equation}
where 
\begin{equation}
u_{0}=p\left(1-\r_{1}^{st}\right)+p'\r_{1}^{st}+q\left(1-\r_{-1}^{st}\right)+q'\r_{-1}^{st}\ge0.\label{eq:u_0}
\end{equation}
However, here we are interested in the model's temporal behavior. In light of Eq. \eqref{eq:rho_bulk_eq_scaled}, self-consistency requires that the leading long-time behavior of $v\left(t\right)$ be
\begin{equation}
v\left(t\right)\asy\frac{\o}{\sqrt{t}},\label{eq:v(t)}
\end{equation}
where $\o$ is an unknown constant.
Since $v\left(t\right)=v_{+}\left(t\right)-v_{-}\left(t\right)$ with
$v_{\pm}\left(t\right)\ge0$, it is reasonable to guess a similar asymptotic behavior for the moving rate $u\left(t\right)$, which we assume to be of the form
\begin{equation}
u\left(t\right)\asy u_{0}+\frac{\eta}{\sqrt{t}}.\label{eq:u(t)}
\end{equation}

\subsubsection{Density profile}

Substituting $\r_{\el}^{\pm}\left(t\right)$ of Eq. \eqref{eq:ansatz}
and $v\left(t\right)$ of Eq. \eqref{eq:v(t)} into Eq. \eqref{eq:rho_bulk_eq} for $\r_{\el}(t)$ and taking the continuum limit as $t\ra\infty$ yields the ordinary differential equation
\begin{equation}
\ddot{\d\r_{\pm}}\left(z_{\pm}\right)=-\frac{2\o+z_{\pm}}{2+u_{0}}\dot{\d\r_{\pm}}\left(z_{\pm}\right),\label{eq:scaled density ODE}
\end{equation}
whose solution is 
\begin{equation}
\d\r_{\pm}\left(z_{\pm}\right)=A_{\pm}+B_{\pm}\text{Erf}\left[\frac{2\o+z_{\pm}}{\sqrt{2\left(2+u_{0}\right)}}\right].\label{eq:density solution}
\end{equation}
The parameters $A_{\pm}$ and $B_{\pm}$ will next be determined
by imposing the appropriate boundary conditions. The first boundary condition is obtained by noting that, at a sufficiently
large distance from the tracer, the density profile decays to $\overline{\r}$, implying that 
\begin{equation}
\d\r_{\pm}\left(z_{\pm}\ra\pm\infty\right)=0.\label{eq:1st BC}
\end{equation}
The second boundary condition is deduced using the known stationary
densities $\lim_{t\ra\infty}\r_{\pm1}\left(t\right)=\r_{\pm1}^{st}$
in Eq. \eqref{eq:extended phase boundary densities}, giving 
\begin{equation}
\d\r_{\pm}\left(z_{\pm}\ra0^{\pm}\right)\asy\r_{\pm1}^{st}-\overline{\r}.\label{eq:2nd BC}
\end{equation}
Accounting for both boundary conditions in Eqs. \eqref{eq:1st BC}
and \eqref{eq:2nd BC}, $\d\r_{\pm}\left(z_{\pm}\right)$ assumes its final form
\begin{equation}
\d\r_{\pm}\left(z_{\pm}\right)=\left(\r_{\pm1}^{st}-\overline{\r}\right)\frac{1\mp\text{Erf}\left[\frac{2\o+z_{\pm}}{\sqrt{2\left(2+u_{0}\right)}}\right]}{1\mp\text{Erf}\left[\sqrt{\frac{2}{2+u_{0}}}\o\right]}.\label{eq:density solution_1}
\end{equation}

The tracer's asymptotic velocity and moving rates are now within reach. Using the bath density deviation $\d\r_{\pm}\left(z_{\pm}\right)$,
we easily compute the long-time behavior of the bath density near
the tracer (i.e. for $\left|\el\right|\sim\ord 1$) as
\begin{equation}
\r_{\pm\el}\left(t\right)\asy\r_{\pm1}^{st}+D_{\pm}\frac{c_{\pm}+\el}{\sqrt{t}},\label{eq:density solution small l}
\end{equation}
where 
\begin{equation}
D_{\pm}=\pm\sqrt{\frac{2}{\pi\left(2+u_{0}\right)}}\frac{\left(\overline{\r}-\r_{\pm1}^{st}\right)e^{-\frac{2\o^{2}}{2+u_{0}}}}{1\mp\text{Erf}\left[\sqrt{\frac{2}{2+u_{0}}}\o\right]}.\label{eq:D_pm}
\end{equation}
The only remaining parameters which must be set to uniquely determine $\r_{\pm\el}\left(t\right)$ are $c_{\pm}$ of Eq. \eqref{eq:ansatz} and $\o$ of Eq. \eqref{eq:v(t)}. The parameters $c_{\pm}$ are set by the boundary Eqs. \eqref{eq:boundary equations} by substituting $\r_{\pm1}\left(t\right)$ and $\r_{\pm2}\left(t\right)$ of Eq. \eqref{eq:density solution small l}. Their explicit expressions are cumbersome and provide little physical insight and are thus not presented. For the extreme parameters they reduce to $c_{+}=c_{-}=0$ and for the bulk parameters that are used to generate the figures in Sec. \ref{sec:Main-Results}, we find $c_{+}\approx 0.336$ and $c_{-}\approx 19.825.$
Finally, a transcendental equation for $\o$ is obtained by demanding that the tracer's velocity $v\left(t\right)=v_{+}-v_{-}$, which is an explicit function of $\r_{\pm1}\left(t\right)$, be consistent with its assumed form $v\asy\frac{\o}{\sqrt{t}}$
of Eq. \eqref{eq:v(t)}. We find
\begin{equation}
\o=\left(1+\frac{u_{0}}{2}\right)\frac{D_{-}-D_{+}}{\r_{+1}^{st}-\r_{-1}^{st}}.\label{eq:d equation}
\end{equation}

With this we conclude our MF analysis of the bath density profile's
evolution and the tracer's velocity in the limit of large $t$. We
next show that, at the extreme points of the extended phase, these MF results remarkably yield the correct asymptotic temporal scaling of the tracer's MSD.

\subsection{A biased random walk with time-dependent rates\label{subsec:MSD}}

While our analysis has thus-far mostly concerned the bath density profile, let us shift our focus to the tracer's dynamics. In general, the tracer may be viewed as a walker which makes right and left moves with some probability. We use the bath density profile to calculate these moving rates and show that at the extreme points these rates decay in time. This, in turn, is responsible for the observed subdiffusive behavior of the tracer, which is recovered even when correlation between its moves are neglected. 

Within this framework we model the tracer's dynamics as a random walker with {\it{time dependent moving rates}}, whose $n^{th}$ discrete step $Z_n$ satisfies
\begin{equation}
Z_{n}=\begin{cases}
+1 & \text{w.p. }P_{n}=\frac{v_{n}^{+}}{u_{n}}\\
-1 & \text{w.p. }Q_{n}=\frac{v_{n}^{-}}{u_{n}}
\end{cases},\label{eq:X_n}
\end{equation}
where 
\begin{equation}
P_{n}+Q_{n}=1\text{ }\text{and}\text{ }P_{n}-Q_{n}=\frac{v_{n}}{u_{n}}.\label{eq:P and Q}
\end{equation} 

Denoting the tracer's position after $N$ steps by $X_{N}=\sum_{n=1}^{N}Z_{n}$ and denoting the difference between the tracer's $n$'th step $Z_{n}$ and its mean value $\left\langle Z_{n}\right\rangle $ by 
\begin{equation}
Y_{n}=Z_{n}-\left\langle Z_{n}\right\rangle,\label{eq:Y_n}
\end{equation}
the tracer's MSD becomes
\begin{equation}
\left\langle \D X_{N}^{2}\right\rangle \equiv\left\langle \left(\sum_{n=1}^{N}Y_{n}\right)^{2}\right\rangle =\sum_{n=1}^{N}\left\langle Y_{n}^{2}\right\rangle +\sum_{m\ne n}\left\langle Y_{m}Y_{n}\right\rangle .\label{eq:MSD Y}
\end{equation}
Neglecting correlations between consecutive steps then
allows factorizing $\left\langle Y_{m}Y_{n}\right\rangle \approx\left\langle Y_{m}\right\rangle \left\langle Y_{n}\right\rangle$,
which vanish since $\left\langle Y_{n}\right\rangle=0$ by definition.
Using $\left\langle Z_{n}^{2}\right\rangle =1$ and $\left\langle Z_{n}\right\rangle =P_{n}-Q_{n}$,
which follow directly from Eq. (\ref{eq:P and Q}), a straightforward
calculation yields
\begin{equation}
\left\langle \D X_{N}^{2}\right\rangle \approx\sum_{n=1}^{N}\left[\left\langle Z_{n}^{2}\right\rangle -\left\langle Z_{n}\right\rangle ^{2}\right]=\sum_{n=1}^{N}\left[1-\left(\frac{v_{n}}{u_{n}}\right)^{2}\right].\label{eq:MSD_discrete}
\end{equation}

To proceed one needs to express the number of moves $N(t)$, that take place during the time interval $(0,t)$, as a function of $t$, for large $t$. Noting that $u(t)dt$ is the
probability that the tracer makes a move during the time interval $dt$, the number of moves $N(t)$ satisfies
\begin{equation}
    \frac{dN}{dt}=u(t)~,
    \label{eq: dN(t)}
\end{equation}
and thus 
\begin{equation}
    N(t)=\int_{0}^{t}u(\tau)d\tau ~.
    \label{eq: N(t)}
\end{equation}
Approximating the sum in Eq. \eqref{eq:MSD_discrete} by an integral, it may be expressed as
\begin{equation}
\left\langle \D X\left(t\right)^{2}\right\rangle \approx\int_{0}^{t}\dif\t u\left(\t\right)\left(1-\left(\frac{v\left(\t\right)}{u\left(\t\right)}\right)^{2}\right)~.
\label{eq:MSD Integral}
\end{equation}

We have seen in Sec. \ref{subsec:MF-Approximation} that at the extreme points of the extended phase, where $u_{0}=0$, the tracer's moving rate and velocity respectively vanish as $u(t)\asy{\eta/\sqrt{t}}$ and $v(t)\asy{\omega/\sqrt{t}}$ at large $t$, for $1\ll t\ll L^2$. On the other hand, performing the analysis described in \cite{miron2019single} at the extreme points shows that in the steady state, i.e. for $t\gg L^2 \gg1$, the tracer's moving rate and velocity respectively vanish as $u^{st}(L) \asy \mu/L$ and $v^{st}(L) \asy \nu/L$. Using these asymptotic limits in Eq. \eqref{eq:MSD Integral} yields the following expression for the MSD

\begin{equation}
\left\langle \D X\left(t\right)^{2}\right\rangle \approx \begin{cases}
\frac{2(\eta^2-\omega^2)}{\eta}\sqrt{t} & t/L^2 \ll 1 \\
\frac{(\mu^2-\nu^2)}{\mu}\frac{1}{L}t & t/L^2 \gg 1
\end{cases},\label{eq:MSD Asymptotic}
\end{equation}
where the parameters $\eta, \omega, \mu$ and $\nu$ can be expressed in terms of the average density $\bar{\r}$ and the dynamical rates $p,q,p'$ and $q'$.
Note that the large $t$ expression is valid only as long as $t \ll L^3$: in a finite system of length $L$, the MSD is bounded from above by $\sim O(L^2)$. At $t\sim O(L^3)$ the MSD reaches this bound and stops growing.

We thus conclude that the MSD  exhibits three distinct types of behavior, separated by two crossover regimes. For $t\ll L^2$ the tracer's dynamics is sub-diffusive, as is the case for SF dynamics. At $t\sim O(L^2)$, a crossover to ordinary diffusive behavior takes place with a diffusion constant $D(L)$ which decays to zero as $\sim 1/L$ at large $L$. Finally, at $t\sim O(L^3)$, the MSD saturates at a value of $\sim O(L^2)$ and reaches a constant value. In the first crossover regime the MSD can be described by a scaling function of the form
\begin{equation}
   \left\langle \D X\left(t\right)^{2}\right\rangle \approx L\chi\left (\frac{\sqrt{t}}{L}\right )~, 
\end{equation}
with 
\begin{equation}
    \chi \left (x \right) \propto \begin{cases}
    x & x\ll 1 \\
    x^2 & x \gg 1
    \end{cases}~.
\end{equation}
This scaling form, obtained within the random walk picture, qualitatively agrees with the simulation results presented in Figs. \ref{MSD}, \ref{MSD_slope} and \ref{MSD_scaled}. There, both the sub-diffusive and diffusive domains of the tracer's dynamics are observed, separated by a crossover regime which takes place at $t\sim O(L^2)$.

An important remark, concerning the neglected correlations between the walker's steps in Eq. \eqref{eq:MSD_discrete}, must be made to correctly frame these results. 
At the extreme points one recovers the correct scaling of the MSD, but not the correct coefficient. To illustrate this point for the extreme parameters used to generate Figs. \ref{MSD_zoom} and \ref{MSD}, one has $\o\approx 0.87$ and $\eta\approx 0.95$, which give $\langle\D X(t)^{2}\rangle\approx 0.28\sqrt{t}$. Yet the fit in Fig. \ref{MSD_zoom}, instead, shows $\langle\D X(t)^{2}\rangle\approx 6.05\sqrt{t}$. Moreover, repeating this analysis in the bulk regime, where the moving rate of the tracer does not vanish in the large-$L$ limit, incorrectly predicts a diffusive scaling, i.e. $\langle\D X(t)^{2}\rangle\pro t$. Both of these discrepancies directly follow from the neglected correlations. At the extreme points, these are less severe due to the tracer's fully biased dynamics, hopping only to the right and exchanging only to the left for $\d=-\d'=1$. As such, the MSD's scaling remains correct, with the correlations merely modifying the prefactor. However, in the bulk regime these correlations are more significant and cannot be ignored. Nevertheless, obtaining the correct sub-diffusive scaling from such a robust MF mechanism provides important physical intuition into such correlated dynamics and is expected to apply to many additional scenarios.

\section{Numerical procedure \label{sec:Numerical procedure}}

We finally present the numerical procedure used to obtain
the simulation results in Sec. \ref{sec:Main-Results}.
Each realization began with drawing the positions of $N=\bar{\r}\left(L-1\right)$ bath particles uniformly over a lattice of $L$ sites with the tracer located at $\ell=0$.
Initial tracer hop and exchange times $\vec{\t}=\left(\t_{p},\t_{q},\t_{p'},\t_{q'}\right)$
were drawn from exponential distributions with the respective
hop and exchange rates $p,q,p'$ and $q'$. For the bath particles,
the Gillespie algorithm was used to draw the initial bath hop time
$\s$ from an exponential distribution with rate $N/2$, accounting
for both right and left hops \citep{gillespie2007stochastic}. The
dynamics was carried out as follows: the smallest of the times $\vec{\t}$ and $\s$ was first determined. If this was $\s$, a bath particle index was next drawn from the $N$ bath particle indices, as well as a random hop direction $\pm1$. The bath particle would then hop to its neighboring right/left site, if the site was vacant. If instead one of the tracer hop times, $\t_{p}$ and $\t_{q}$, was the smallest, the tracer would hop to the right/left neighboring site, again, if the site was vacant. If one of the tracer exchange times, $\t_{p'}$
and $\t_{q'}$, was the smallest, the tracer would exchange places
with a bath particle to its right/left if a bath particle was present at that site. Following any of the above scenarios, a new time was drawn and the remaining times were updated.

\section{Conclusions\label{sec:Conclusions}}

In this paper we have studied the dynamics of a 1D driven tracer moving through a geometrically-confined quiescent bath. For the non-driven tracer, when strong geometric confinement prevents particles from overtaking one another, one famously recovers the sub-diffusive behavior which characterizes single-file dynamics.
Our main result is that for the driven tracer, this sub-diffusive behavior remarkably persists even when the degree of confinement is reduced and overtaking is allowed. This stands in contrast with the non-driven tracer dynamics, which are well-known to become diffusive at \textit{any} finite overtaking rate.

In \cite{miron2019single}, the model's steady state was studied and found to exhibit an extended phase, where the bath density profile extends throughout the entire system and the tracer's velocity vanishes in the thermodynamic limit. 
Here we have focused on the model's dynamical properties in this phase.
Using the MF approximation, we have computed the bath density profile's temporal evolution $\r_\el(t)$, as viewed from the tracer's reference frame. This was shown to approach a scaling function $\r(\el/\xi)$ with a characteristic length $\xi$ which grows in time as $\sim\sqrt{t}$. Moreover, we have shown that the tracer's velocity $v(t)$ asymptotically scales as $\sim 1/\sqrt{t}$ for large $t$.
These results have allowed us to model the tracer's dynamics as a biased random walk with time-dependent rates and compute its MSD at specific regions of the extended phase. Using this approach, we show that the tracer's dynamics remains sub-diffusive with $\left\langle \D X\left(t\right)^{2}\right\rangle \sim\sqrt{t}$, even for finite overtaking rates, and confirm this picture using extensive numerical simulations.

\section{Acknowledgments\label{sec:Acknowledgments}}

We thank Julien Cividini, Bertrand Lacroix-A-Chez-Toine and Harald A Posch for discussions and suggestions. This work was supported by a research grant from the Center of Scientific Excellence at the Weizmann Institute of Science.

\end{document}